\documentclass[10pt,journal,compsoc]{IEEEtran}
\newif\ifpeerreview
\peerreviewfalse

\usepackage[ruled,vlined,linesnumbered]{algorithm2e}
\usepackage{amsmath}

\DeclareMathOperator*{\argmin}{arg\,min}
\newcommand{\norm}[1]{\left\lVert#1\right\rVert}
\usepackage{array, makecell}

\usepackage[nocompress]{cite}
\usepackage{url}
\usepackage{amsmath,amssymb,graphicx}
\usepackage{xcolor}
\usepackage[switch]{lineno}
\usepackage{capt-of}

\title{Automatic calibration of time of flight based non-line-of-sight reconstruction}

\author{Subhash Chandra Sadhu,
        Abhishek Singh,
        Tomohiro Maeda,
        Tristan Swedish,
        Ryan Kim,
        Lagnojita Sinha,
        and Ramesh Raskar
}

\begin{document}

\IEEEtitleabstractindextext{%
\begin{abstract}
Time of flight based Non-line-of-sight (NLOS) imaging approaches require precise calibration of illumination and detector positions on the visible scene to produce reasonable results. If this calibration error is sufficiently high, reconstruction can fail entirely without any indication to the user. In this work, we highlight the necessity of building autocalibration into NLOS reconstruction in order to handle mis-calibration. We propose a forward model of NLOS measurements that is differentiable with respect to both, the hidden scene albedo, and virtual illumination and detector positions. With only a mean squared error loss and no regularization, our model enables joint reconstruction and recovery of calibration parameters by minimizing the measurement residual using gradient descent. We demonstrate our method is able to produce robust reconstructions using simulated and real data where the calibration error applied causes other state of the art algorithms to fail.
\end{abstract}

}

\twocolumn[{%
\renewcommand\twocolumn[1][]{#1}%
\maketitle
\begin{center}
\centering
\includegraphics[width=0.9\linewidth]{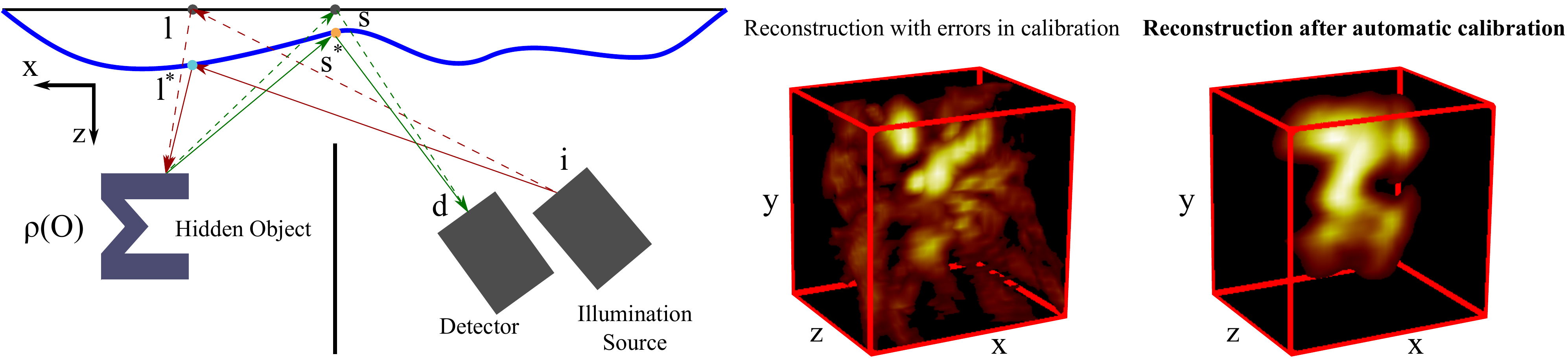}
\captionof{figure}{\label{fig:teaser} \textbf{We propose automatic calibration framework for non-line-of-sight (NLOS) imaging to demonstrate 3D reconstruction around a corner with errors in the estimation of the scanning and detection points on the visible surface.} Existing NLOS reconstruction techniques are susceptible to errors in the estimation of the laser illumination scan $\mathbf{l}$ and detector focus spot $\mathbf{s}$ locations. The proposed framework performs automatic calibration on transient light measurements to robustly recover the hidden object, ``Sigma'' with (3cm) of error applied to illumination scan positions.}

\end{center}%
}]

  \section{Introduction}\label{sec:introduction}
%
%
%
%
\IEEEPARstart{A}{dvances} in high-speed detectors, light sources, and reconstruction algorithms based on solving the inverse light transport problem have made it possible to locate and reconstruct objects around corners~\cite{maeda2019recent} using Time of Flight (ToF) based techniques. When light scatters off diffuse surfaces visible to both the observer and hidden scene, information about the direction light came from before scattering is lost. This makes it challenging to determine what is around corners directly. By using knowledge of the geometry of the scene, and by sending very short pulses of light into the scene to obtain light transients using an ultra-fast sensor, 3D reconstruction of the hidden scene can be achieved. Such computational imaging modalities model light transport around corners to reconstruct hidden objects \cite{Velten12,OToole:2018:ConfocalNLOS}. NLOS imaging has the potential to expand the capability of the current sensing technology for useful applications such as autonomous navigation and medical endoscopy.

\begin{figure}[t]
\begin{center}
  \includegraphics[width=0.9\linewidth]{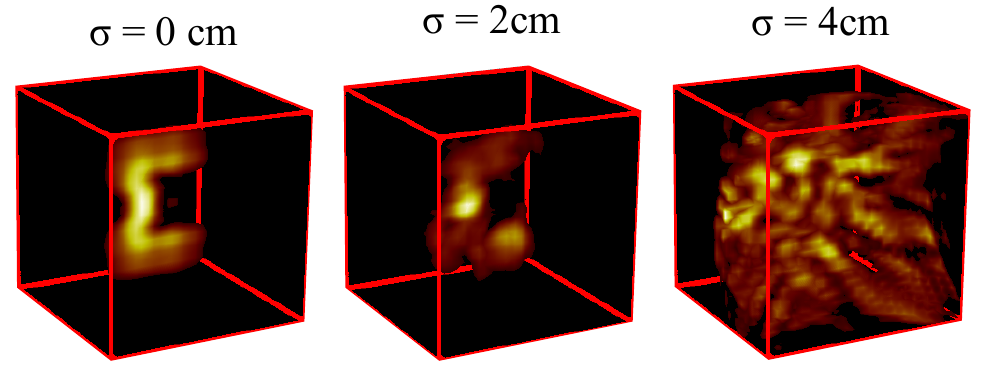}
\end{center}
  \caption{\textbf{NLOS imaging requires accurate estimation of the scanning and detection positions.} Phasor field reconstruction~\cite{liu2019phasor_nlos} is performed on simulated measurement. Reconstruction uses scanning and detection location with errors that follow zero-mean gaussian distribution with standard deviation $\sigma$. The reconstruction quality degrades with small errors, and ``Sigma'' reconstruction is not recognizable at $\sigma = 4cm$.}
\label{fig:noise_sensitivity}
\end{figure}

Following the first 3D reconstruction around a corner by Velten et al.~\cite{Velten12}, various algorithms for Non-Line-of-Sight imaging (NLOS) with ToF measurement have been proposed for faster and more robust reconstruction\cite{lindell2019wave,OToole:2018:ConfocalNLOS, liu2019phasor_nlos}. These methods assume precise estimation of the scanning and detection spot locations, requiring careful calibration of each corner scene. As demonstrated in Fig.~\ref{fig:noise_sensitivity}, even small errors in the calibration parameters such as scanning positions degrade the reconstruction quality. The necessity of precise calibration may limit NLOS imaging in practice, as the calibration of the corner scene often requires a time-consuming and error-prone procedure.

In contrast to past works that have focused on reconstruction given careful calibration, we propose a NLOS reconstruction framework that is robust to errors in calibration, specifically the estimate of scanning and detection positions in the visible scene. Our framework exploits a differentiable forward model that computes the gradient of the estimated hidden volume and calibration parameters with respect to the residual error between model outputs and measurements. The ability to optimize over the calibration parameters in the reconstruction enables automatic calibration from the measurements alone.

\par
\subsection{Contributions}

The main contributions of our work are:
\begin{itemize}
\item Modification to the traditional NLOS forward model, making it differentiable with respect to scanning and detector position along the visible scene surface.
\item Framework for automatic calibration of time of flight based non-line-of-sight measurement and simultaneous reconstruction of the hidden scene using gradient descent.
\item Demonstration of recovery of a hidden object from perturbed calibration parameters for both simulated and real data.
\end{itemize}

Our proposed automatic calibration technique refines estimates of laser scanning and detection points throughout the reconstruction process. While previously proposed algorithms for NLOS imaging require precise point locations on the visible surface, our framework can handle and correct inaccurate estimates of the laser and detection points without any regularization on the reconstruction. This is an important step towards practical applications of NLOS imaging, where visible surfaces may have irregularities that complicate calibration using current methods (e.g., fitting a planar surface using a calibration chart). For the first time, we demonstrate NLOS imaging with imprecise estimates of the laser scanning and detector focus spots.

\subsection{Scope}
This paper takes the first step towards implementing NLOS imaging ``in the wild'' by demonstrating hidden scene reconstructions even with errors in calibration parameters. We assume that the laser scanning positions and detection positions are partially known, with errors sufficiently small that a good solution can be found using gradient descent. The analysis of the convergence for the degree of errors in calibration is not within the scope of this paper. We note that there are parameters such as the bidirectional reflectance distribution function (BRDF) of the visible relay surface and the hidden object that we do not consider. In this paper, we focus on NLOS imaging with an inadequately precise measurement of the scene geometry.

\section{Related Work}

\subsection{Non-Line-of-Sight Imaging}
Non-line-of-sight (NLOS) imaging was first demonstrated using time-of-flight (ToF) measurements~\cite{Raskar5DT,Kirmani2009,Velten12}. The algorithm used for the first 3D reconstruction was filtered backprojection~\cite{Velten12}. However, filtered backprojection does not provide an exact solution for the light transport equation which gives a blurred reconstruction. Since then, more robust algorithms have been proposed to address this issue. Regularized linear inversion incorporates priors in reconstruction to increase the reconstruction quality~\cite{Gupta12, Heide:2019:OcclusionNLOS, Kadambi16, Ahn_2019_ICCV}. Confocal NLOS setup~\cite{OToole:2018:ConfocalNLOS} provides a closed-form solution to the inverse problem of reconstruction, and can be combined with a wave-imaging based technique~\cite{Lindell:2019:Wave}. More recently, wave-based approach~\cite{liu2019phasor_nlos} and geometrical approach~\cite{Tsai17,Xin:19} have been demonstrated to perform robust NLOS imaging. The wave based approach is also called phasor field reconstruction. We employ phasor field reconstruction in this paper to baseline and compare reconstruction results.

Information other than ToF is also useful to see around corners. For instance, shadows cast by an occluding wall and occluders in the hidden scene provide cues on the hidden object for tracking~\cite{Bouman17} and also 2D reconstruction~\cite{Saunders2019Periscopy} employing standard RGB camera hardware. Data-driven approaches~\cite{Tancik2018FlashPF} have been used to learn to exploit intensity measurements other than shadows. Speckle patterns from coherence of light has been used to perform 2D reconstruction~\cite{Katz14} and tracking~\cite{Smith_2018_CVPR} at a microscopic scale.

While past work on ToF-based NLOS imaging has aimed for robustness and efficiency of reconstruction, all those algorithms require a well-calibrated scene. In contrast, our work aims to handle errors in scene calibrations by correcting such errors in the reconstruction process. Here, auto-calibration is enabled by making the imaging system's mathematical model differentiable with the calibration parameters.

\begin{figure*}[!t]
\centering
\includegraphics[width=0.9\linewidth]{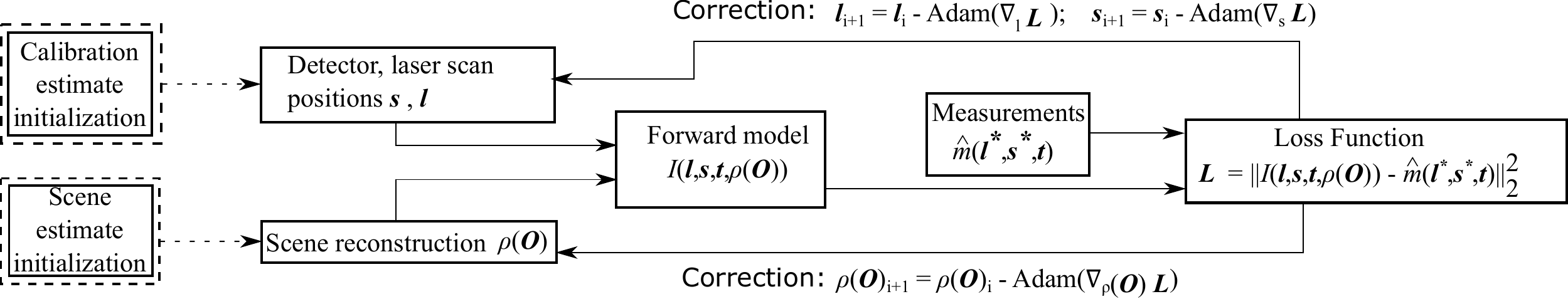}
\caption{\textbf{Autocalibration block diagram.} Our method is initialized with a guess of the calibration parameters and reconstructs the hidden scene and updates calibration parameters while ensuring consistency with light transient measurements.}
\label{blk_diagram}
\end{figure*}

\subsection{Differentiable Imaging Systems}
Auto-differentiation (AD) has become an important tool for optimizing a system catered to a given objective. If an imaging system can be differentiated with respect to some parameters, then gradient based optimization techniques can be used to solve for the parameters that minimize an objective function. Popularized by deep learning frameworks such as PyTorch and Tensorflow, AD seeks to automatically determine the gradient of general computer programs. For an in-depth review of AD within the context of deep learning frameworks see~\cite{baydin2018automatic}. 

Recently, there have been proposals for imaging system design using differentiable models, such as Microscope design for improved classification~\cite{chaware2019towards, muthumbi2019learned}, joint optimization of convolutional neural networks and phase-masks for improved depth-of-field~\cite{Elmalem:18}, designing color sensor masks for digital image sensors that improve demosaicing~\cite{chakrabarti2016learning}, extended depth-of-field and super resolution for natural images~\cite{sitzmann2018end}, and designing phase masks for monocular depth estimation~\cite{wu2019phasecam3d}. Recently, AD has been proposed as a solution to enable differentiable path tracing for full light transport~\cite{NimierDavidVicini2019Mitsuba2}.

Rather than using AD to solve for an imaging design, our work employs AD to jointly solve for the reconstruction and calibration parameters of our NLOS imaging system.

\subsection{Auto-Calibration}
Refinement of calibration parameters for imaging has been well studied for computer visions applications. Auto-calibration in vision is often referred to as ``bundle adjustment''~\cite{Triggs2000}, which jointly optimizes for the estimate of 3D structure of the scene and camera viewpoints. Bundle adjustment is the underlying problem considered in photogrammetry~\cite{Granshaw2006}, structure from motion~\cite{Mouragnon2009, Bartoli2005}, and simultaneous localization and mapping (SLAM)~\cite{Konolige2008}. Automatic calibration is also used for medical imaging such as computed tomography~\cite{Wein2011,Ladikos2012} and ultrasound imaging~\cite{Blackall2000AnIR}, and ptychography~\cite{Eckert:18, Thibault2009}, where the quality of image reconstruction is dependent on precise calibration.

We introduce an auto-calibration framework to NLOS imaging for the first time. Since NLOS imaging is sensitive to the quality of scene specific calibration, the introduction of such an auto-calibration framework is an important step towards practical NLOS imaging.

\section{Proposed Method}

\subsection{Light Transport}
Fig.~\ref{fig:teaser} illustrates the light transport for NLOS imaging. A pulsed laser illuminates a point $\mathbf{l}$ on the visible surface where photons scatter isotropically, reach the hidden object $\mathbf{O}$, and scatter again at a point $\mathbf{s}$ on the visible surface. Denoting the illumination source and detector positions as $\mathbf{i}$ and $\mathbf{d}$, and the surface albedo of the hidden scene as $\rho(\mathbf{o})$, the measurement of the three bounce photons depicted in Fig.~\ref{fig:teaser} can be expressed as follows:

\begin{equation}
    d_{\text{total}} = \norm{\mathbf{l} - \mathbf{o}} + \norm{\mathbf{s}-\mathbf{o}} + \norm{\mathbf{i} - \mathbf{l}} + \norm{\mathbf{d} - \mathbf{s}}
\end{equation}

\begin{equation}
    m = \int_\mathbf{O} \rho(\mathbf{o}) \frac{\delta(ct -  d_{\text{total}})}{\norm{\mathbf{l} - \mathbf{o}}^2 \norm{\mathbf{s} - \mathbf{o}}^2} d\mathbf{o}
\label{eqn_light_transport}
\end{equation}

where $\mathbf{i}$ and $\mathbf{d}$ are fixed for a given imaging setup and $d_{\text{total}}$ is the path length a photon emitted from the light source $i$, and scattered from the hidden scene $o$ must travel to reach the detector $d$. The dirac delta function $\delta(\cdot)$ comes from the time-of-flight measurement, and the denominator represents the intensity fall off as the square of the radial distance due to diffuse scattering at the visible and the hidden surface.

\subsection{Differentiable Model for reconstruction}
Instruments only measure time in discrete time bins and we must describe a finite set of positions in the hidden scene to make Equation~\ref{eqn_light_transport} computable. Thus, our forward model is redefined for a finite set of $N_O$ voxels.

\begin{equation}
    \hat{m} = \sum_\mathbf{i}^{N_{O}} \rho(\mathbf{o_i}) \frac{\Pi\bigg( { k - \frac{d_{\text{total, i}}}{c\Delta t}} \bigg)}{\norm{\mathbf{l} - \mathbf{o_i}}^2 \norm{\mathbf{s} - \mathbf{o_i}}^2}
\label{eq:discretized_forward_model}
\end{equation}
where $k$ is the index of the time bin, $\Delta t$ is the width of the time bin, and $\Pi(x)$ denotes a rectangle function, $\Pi(x)$ = 1 for $0 < x < 1$ and $\Pi(x)$ = 0 otherwise. 


Autocalibration requires the forward physics model to be differentiable with respect to $\mathbf{s}$, $\mathbf{l}$, and $\rho(\mathbf{o})$. If we examine Eq.~\ref{eq:discretized_forward_model}, the window operator $\mathbf{\Pi}$ models the time binning that occurs in the photon counting instrument as impulses of light reach the detector, which are assigned to discrete time bins of width $\Delta t$ by the photon counting instrument. We are interested in updating $\mathbf{l}$ and $\mathbf{s}$ in order to perform autocalibration. The problem is that the partial derivative of all parameters that pass through $\mathbf{\Pi}$ will be zero, and thus any gradients used to update $\mathbf{l}$ and $\mathbf{s}$ will also be zero if we use Eq.~\ref{eq:discretized_forward_model} directly. To make the forward model differentiable, we model each impulse as a Gaussian so that the partial derivative with respect to $\mathbf{l}$ and $\mathbf{s}$ are not zero.

\begin{equation}
    I(\mathbf{l}, \mathbf{s}, k; \rho(\mathbf{O})) = 
    \sum_\mathbf{i}^{N_{O}} \rho(\mathbf{o_i}) \frac{\mathbf{e}^{-(c\Delta t k - d_{\text{total, i}})^2/\sigma^2}}{\norm{\mathbf{l} - \mathbf{o_i}}^2 \norm{\mathbf{s} - \mathbf{o_i}}^2}
\label{eq:differentiable_forward_model}
\end{equation}

where $\mathbf{\sigma}$ parameterizes the spread of the gaussian function that replaces the impulses. The spread of the gaussian ensures that an impulse in a given time bin, indexed with $k$, actually affects the evaluation of the forward model in time bin, $k+1$. Each accumulated time bin is affected to a small extent by all the time bins.

Our goal is to estimate the illumination $\mathbf{l}$ and detection point $\mathbf{s}$, and reconstruct the hidden volume $\mathbf{O}$. This can be achieved by solving the following optimization problem:
\begin{equation}
    \argmin_{\mathbf{l}, \mathbf{s}, \rho(\mathbf{O})} \sum_\mathbf{s} \sum_\mathbf{l} \sum_k \norm{ I - \hat{m}(\mathbf{s}^*, \mathbf{l}^*, k)}_2^2
\label{eq:cost_function}
\end{equation}
where $\hat{m}(\mathbf{s}^*, \mathbf{l}^*, k)$ is the measurement of the three-bounce photon and $\mathbf{s}^*$ and $\mathbf{l}^*$ denotes the true position of the illumination and detection points. For our experiments, the illumination spots are scanned and the detection spot is fixed on the relay surface at a single location. However, the sum can be taken over detection points as well if multiple detectors are used in an experiment.

In actual implementation, we assume that we have a rough initial estimate of the illumination and detector spot positions. The initialization of the albedo estimate can be made using phasor field backprojection or other reconstruction algorithms using the rough illumination and detector spot positions.

\begin{algorithm}
\SetAlgoLined
 \textbf{Input:} \\ estimated\_laser\_scan\_positions $\mathbf{l}$, \\ estimated\_detector\_focus\_position $\mathbf{s}$, \\ measurement\_real $\mathbf{\hat{m}}$\;
 \textbf{Initialize:} $\rho(\mathbf{O})$ = phasor\_field\_back\_projection($\mathbf{l}$,$\mathbf{s}$,$\mathbf{m}$)\;
 \For {total\_iterations}{
  \For{calib\_iterations} {
  // Calibration iteration\;
  $\mathbf{I}$ = forward\_model($\mathbf{l}$, $\mathbf{s}$, $\rho(\mathbf{O})$)\;
  loss = loss\_function($\mathbf{I}$, $\mathbf{\hat{m}}$)\;
  // Gradient update (Adam)\;
  $\mathbf{l}$, $\mathbf{s}$ = optimizer\_step(loss)\;
  }
  \For{reconstruction\_iterations}{
  // Reconstruction iteration\;
  $\mathbf{I}$ = forward\_model($\mathbf{l}$, $\mathbf{s}$, $\rho(\mathbf{O})$)\;
  loss = loss\_function($\mathbf{I}$, $\mathbf{\hat{m}}$)\;
  // Gradient update (Adam)\;
  $\rho(\mathbf{O})$ = optimizer\_step(loss)\;
  }
 
 }
 \caption{Procedure: AUTOCAL($\mathbf{l}$,$\mathbf{s}$,$\mathbf{0}$,$\mathbf{\hat{m}}$ )}
 \label{algo}
\end{algorithm}

\section{Implementation}
\subsection{Experimental Setup}
\begin{figure}[!t]
\centering
\center{\includegraphics[scale=0.06]{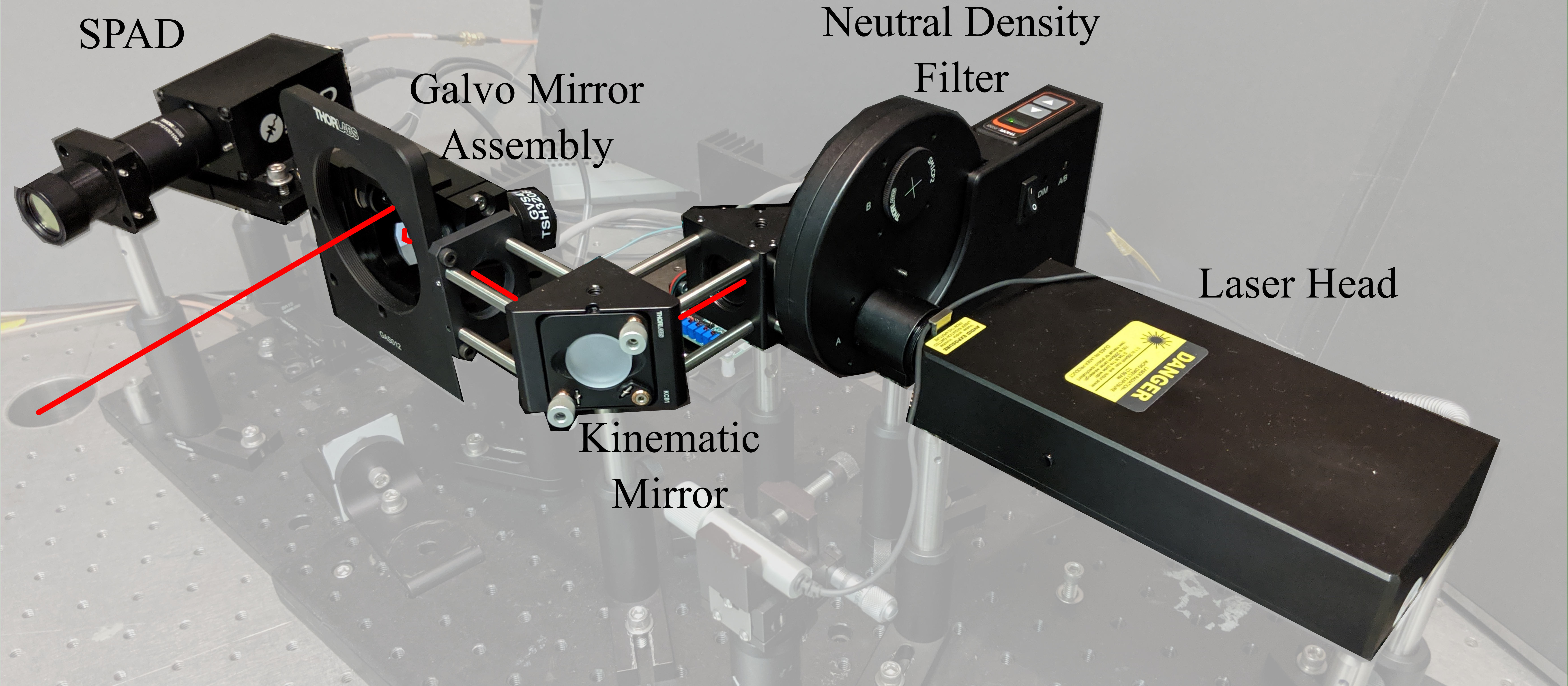}}
\caption{\textbf{Experimental setup:} A collimated, pulsed laser source, kinematic mirrors for beam alignment, articulated galvo mirror system, and single pixel SPAD detector with focussing optics and an optical filter. Setup measures Time of Flight of light by measuring the delay between laser pulse emission, and reflected photon detection by the SPAD. This time delay is measured at a resolution of 1ps or higher by an instrument called Time Correlated Single Photon Counter (TCSPC) which receives electrical signals from the SPAD and the laser. TCSPC is not shown in the setup photo.}
\label{Exp_setup}
\end{figure}

\begin{figure*}[t]
\begin{center}
  \includegraphics[width=0.8\linewidth]{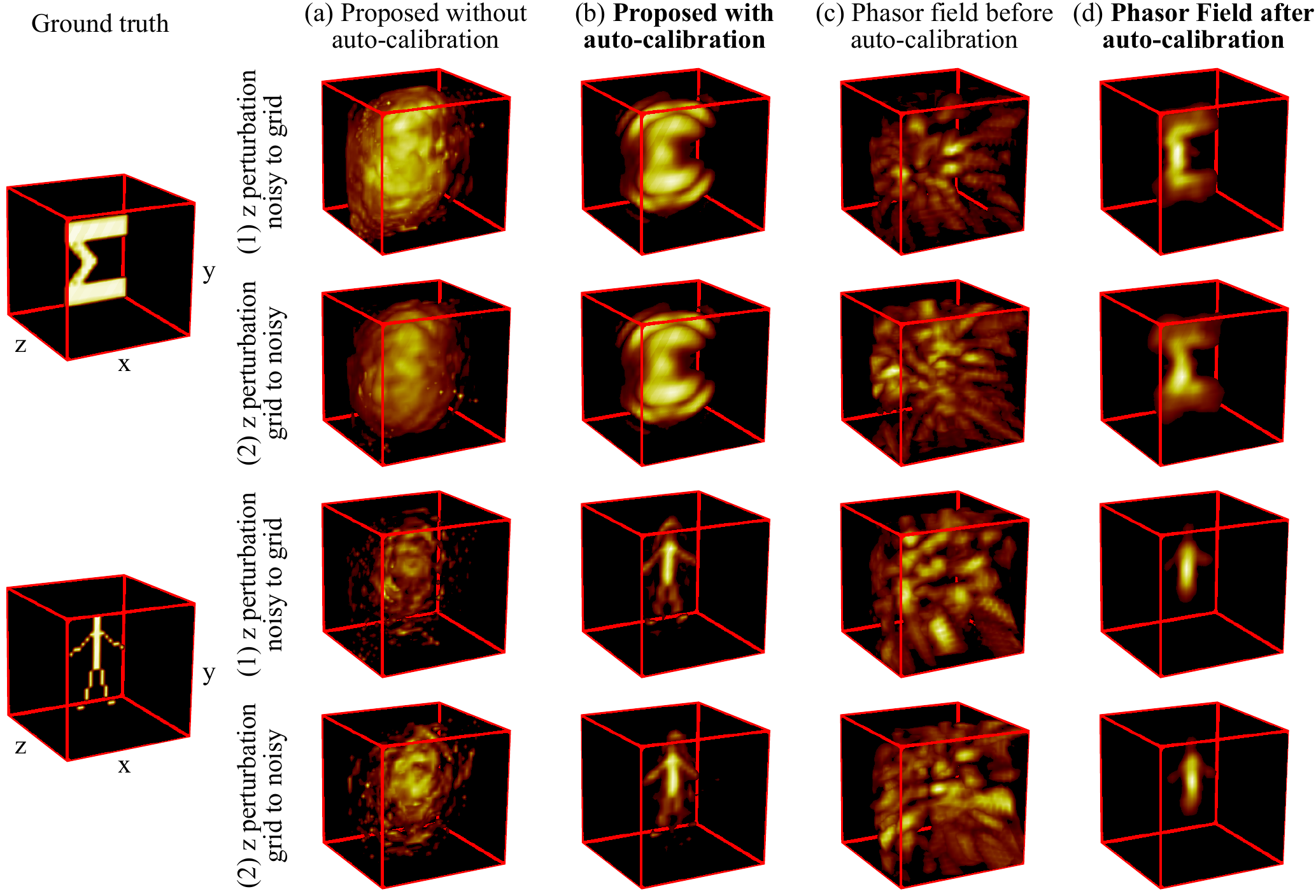}
\end{center}
  \caption{\textbf{The proposed automatic calibration framework reconstructs the hidden object despite errors in calibration parameter estimation for simulated data.} (a) Reconstruction using the proposed framework without correcting the estimate of the scanning and detection points. Optimization on the objective function fails without automatic calibration (b) Proposed method with automatic calibration recovers the hidden object. (c) Phasor field reconstruction~\cite{liu2019phasor_nlos} fails with inaccurate calibration. (d) Phasor field reconstruction recovers the hidden scene with auto-calibrated parameters. (1) Ground truth laser scan positions lie on a flat plane, initialization is not on a flat plane. (2) Ground truth laser scan positions are not on a flat plane, initialization is on a flat plane.}
\label{fig:simulation_results}
\end{figure*}
The optical hardware measures the Time of Flight for light propagation as the system performs measurements. The experiment is represented in Fig.~\ref{fig:teaser} as a simplified source and detector. In the illumination path, the source is a mode locked laser that generates pulses with average power 150mW, FWHM pulse width of \char`\~ 90fs, repetition rate of 10 MHZ, and a center wavelength of 780nm (Carmel, Calmar, USA). The laser's output goes into free space and it is coupled to free space with a laser head. Fig. \ref{Exp_setup} shows the hardware setup. Light exits from the laser head and passes through a motorized neutral density filter wheel which can control its attenuation (FW102C, Thorlabs, USA). Light is then reflected by two kinematic right angled mirrors which are adjusted to center and align the light. Light is then deflected by a pair of programmable Scanning Galvo mirrors (GVS412, Thorlabs, USA). The angle of the mirrors can be controlled to direct the collimated light pulse onto the relay surface at a scanning position of choice.

In the detection path, light is optically filtered by a laser-line band pass filter (FL780-10, ThorLabs, USA, $780 \pm 10$ nm) followed by a lens with a focal length of 50mm and an aperture of 25.4mm. Light coming from the relay surface is focussed on the active area of a Single photon avalanche diode (SPAD) (PDM, MPD). The SPAD is connected to a time correlated single photon counting device (Hydraharp, Picoquant, Germany) which records the time transient measurements by time stamping every arriving photon with a time resolution of 1ps or higher. It achieves time correlation between the received light and transmitted light using a synchronization signal from the mode locked laser. The measured instrument response function consists of a FWHM timing jitter of approximately 40 ps, and we use the instrument with time binning of 16 ps.

\subsection{Software Implementation}
We implement the forward model in Equation~\ref{eq:differentiable_forward_model} with PyTorch~\cite{pytorch}. The optimization procedure is summarized in Alg~\ref{algo}. We use Adam optimizer~\cite{adam} for gradient descent with learning rate $0.01$ and default momentum parameters. We parameterize the scene with scanning and detection positions, temporal resolution, system impulse response, system time offset, and the hidden volume voxel positions. To make the memory requirements of this algorithm tractable, we do not perform gradient descent steps using all the scanning spots in every iteration. This approach is a form of batched stochastic gradient descent where we randomly sample the laser scan positions without replacement to compute approximate gradients across iterations.

\begin{figure}[t]
\begin{center}
  \includegraphics[width=0.8\linewidth]{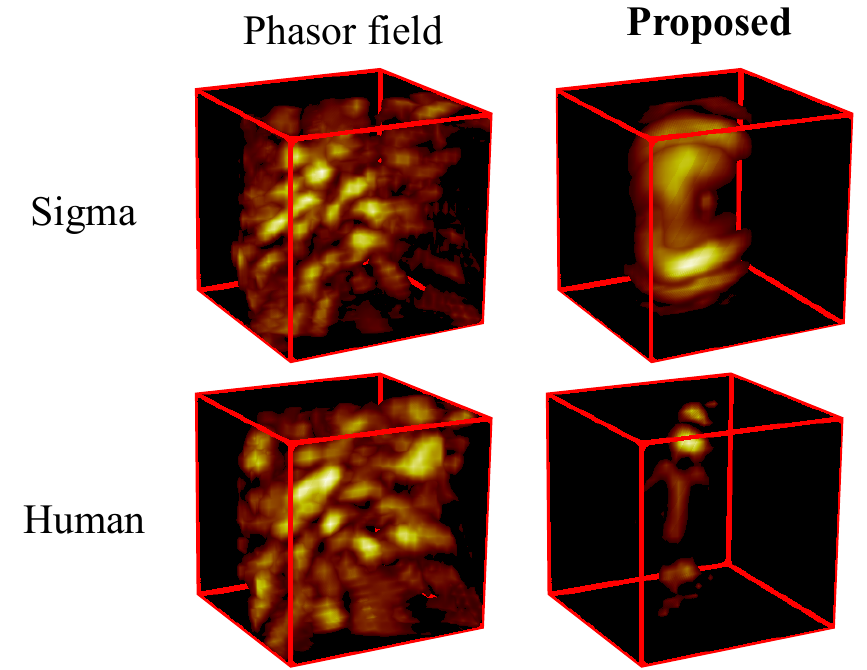}
\end{center}
  \caption{\textbf{Reconstruction on the simulated data with perturbation on x, y, z direction.} The proposed method demonstrates robustness to noise on the estimated scanning and detection points in all x, y, z direction.}
\label{fig:simulation_results_xyz}
\end{figure}

\section{Experimental Validation}
We demonstrate reconstruction of various hidden scenes when the provided scanning positions and detector positions are inaccurate. We validate our algorithm using both simulated and experimental data.

\subsection{Simulation}
In order to demonstrate the feasibility of our autocalibration method, we generate simulated data by synthesizing the measurement of three-bounce photons using our forward physics model with correct calibration parameters. We validate autocalibration by testing under three different miscalibration scenarios:
\begin{enumerate}
\item Synthesize the measurement with ground truth scanning positions in grid, and add random Gaussian noise in z direction to the ground truth of scanning and detection points as an initial estimate.
\item Use grid scanning position as an initial estimate, and synthesize measurements with the scanning and detection positions with Gaussian noise in z direction.
\item Repeat configuration 1, except that the noise is added to all directions (x,y,z).
\end{enumerate}

The first scenario applies to the case where the planar surface is scanned, but the estimation of the scanning points are not accurate. The second scenario considers the case where the surface has small bumps and scan position estimates are as if the surface is planar. The last scenario demonstrates that our method is robust to the errors in scanning and detection point estimation in any direction.

\begin{table}
\setlength{\arrayrulewidth}{.3mm}
\setlength{\tabcolsep}{5pt}
\renewcommand{\arraystretch}{3}
\fontsize{7}{7}\selectfont
\begin{center}
\caption{\textbf{Root mean square error (RMSE) of the scanning position estimate in z direction.} Automatic calibration reduces the errors in scan location estimation.} 
\begin{tabular}{ c|c|c|c|  }
\cline{2-4}
 & \makecell{(1) z perturbation \\ noisy to grid} &  \makecell{(2) z perturbation \\ grid to noisy} &\makecell{(3)xyz perturbation\\ noisy to grid} \\ 
\hline
\multicolumn{1}{|c|}{\makecell{Sigma \\ (Simulation)}} &  \makecell{4.9 cm (initial) \\ \textbf{2.2 cm (recovered)}} &  \makecell{4.8 cm  \\ \textbf{1.8 cm} }  & \makecell{4.9 cm  \\ \textbf{3.0 cm} }\\
\hline
\multicolumn{1}{|c|}{\makecell{Human \\ (Simulation)}} & \makecell{4.9 cm  \\ \textbf{1.5cm} }   &  \makecell{4.9 cm  \\ \textbf{2.0 cm} }  &  \makecell{4.9 cm  \\ \textbf{3.4 cm} } \\
\hline
\end{tabular}
\end{center}
\label{tab:RMSE}
\end{table}

\begin{figure*}[t]
\begin{center}
  \includegraphics[width=0.9\linewidth]{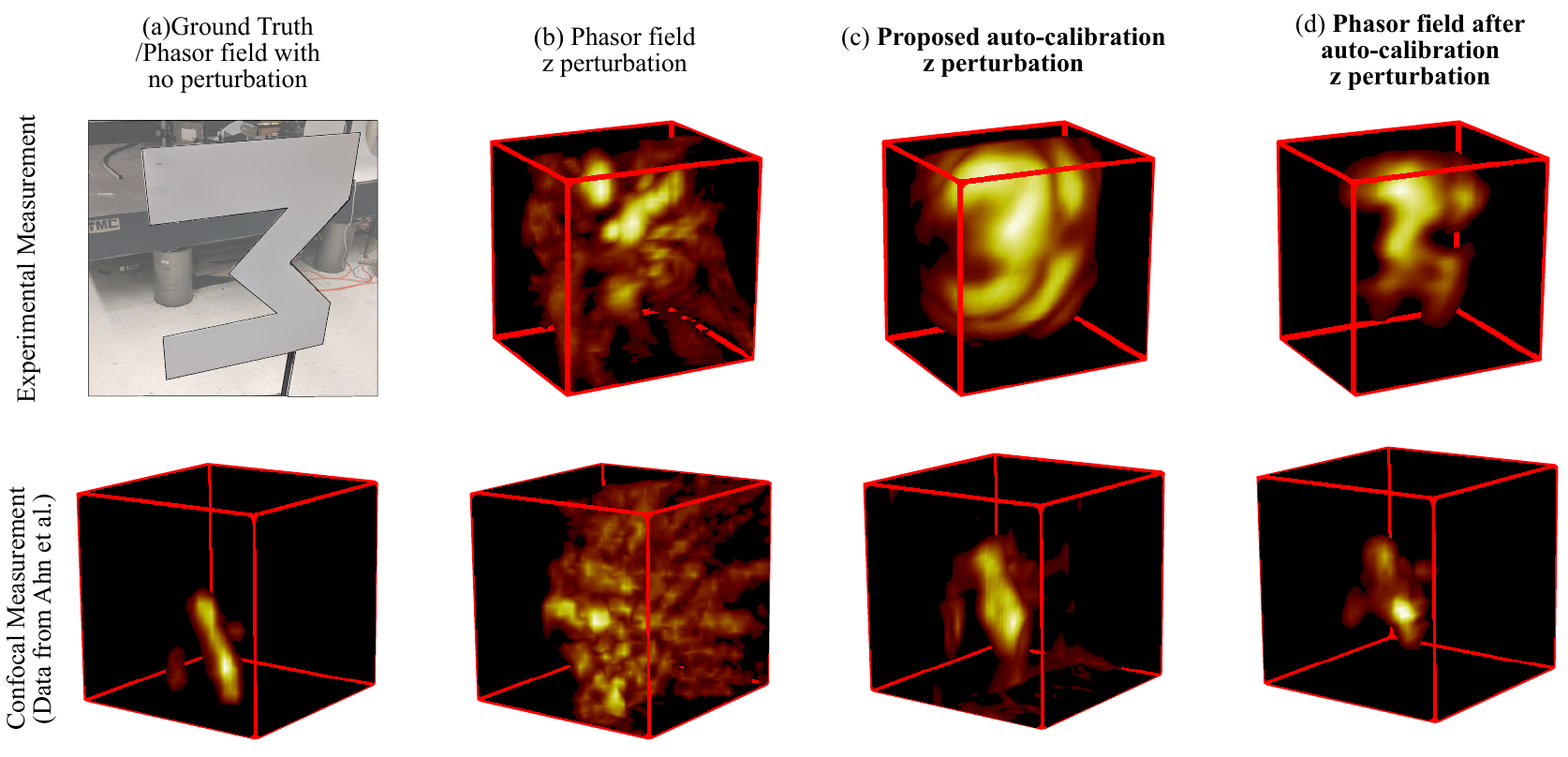}
\end{center}
  \caption{\textbf{Experimental validation of autocalibration.} The top row corresponds measurements made with our experimental system, and the bottom row shows reconstruction from an external dataset~\cite{Ahn_2019_ICCV} using a confocal measurement configuration.}
\label{fig:experiment_results}
\end{figure*}

Our simulation results do not have the potential model mismatch that can occur between our forward model and experimental data. Model mismatch could occur due to various sources of non-idealities such as jitter in SPAD, TCSPC or dark counts in SPAD. Another reason for model mismatch is that the output of the forward model considers only the part of the scene that is inside the reconstruction volume. Even in absence of the mentioned non-idealities, the real data would contain information of reflections from parts of the scene outside the reconstruction volume

\subsubsection*{Results}
We simulate the measurement with laser scan positions over a $32 \times 32$ grid across a square of dimensions $1.28m \times 1.28m$. The random noise added to the estimate of the scan and detection positions follows a zero-mean Gaussian distribution with a standard deviation of $5cm$. We reconstruct a volume defined by $32 \times 32 \times 32$ voxels, where each voxel is a cube of dimensions $4 cm \times 4 cm \times 4cm$. 

In Fig.~\ref{fig:experiment_results}, we reconstruct the hidden objects with (a) phasor field with initial estimates of scanning and detection positions, (b) the proposed framework without corrections on the estimates of scanning and detection position, (c) the proposed framework with auto-calibration, and (d) phasor field reconstruction on auto-calibrated parameters. We alternate the gradient descent on voxel and calibration parameters with 20 iterations each, and repeat the alternation for 4 times for the simulated measurement with ``Sigma'' in the hidden scene, and 7 times for the ``human'' in the hidden scene. 


Fig.~\ref{fig:simulation_results} summarizes the reconstruction results with existing technique~\cite{liu2019phasor_nlos} and our framework on the estimate of scanning and detection positions with errors. The proposed approach successfully recovers the hidden object despite calibration errors. Furthermore, phasor field reconstruction can be performed after the automatic calibration. While we update the calibration parameters only in z direction, this reconstruction demonstrates robustness to errors in x and y directions (Fig.~\ref{fig:simulation_results_xyz}). Table~\ref{tab:RMSE} summarizes error reduction in laser scan positions by the automatic calibration framework.

\subsection{Real Data}
We validate our framework using measurements we collect using the imaging setup described in Section~\ref{Exp_setup} and also with confocal NLOS measurement that is publicly available. Our experimental system is calibrated such that we know the ground truth locations of the scanning positions. We then introduce additive white Gaussian noise to the z-axis of these calibration parameters.

\subsubsection*{Results with Experimental Data}
We capture time-of-flight measurements in NLOS setting with the imaging system described in Section~\ref{Exp_setup}. The relay wall is scanned over a rectangle of dimensions $1.9m \times 1.6m$, from which we capture 1100 sets of light transient measurements. We add zero mean gaussian noise with a standard deviation of 3cm along the z-axis on the calibrated scan and detection locations. We alternate the gradient descent on voxel and calibration parameters with 20 iterations each, and repeat the alternation for 3 times to recover the hidden object.

The added noise in the calibration parameters makes phasor field reconstruction fail, while our framework demonstrates robust recovery of the ``Sigma'' target. After automatic calibration, phasor field results improves with the refined estimation of the scanning and detection locations (Fig.~\ref{fig:experiment_results} Top).

\subsubsection*{Confocal Measurements from Public Dataset}
We also demonstrate our method on real confocal measurements using the dataset made publicly available from~\cite{Ahn_2019_ICCV} (Fig.~\ref{fig:experiment_results} Bottom). We first preprocess the ``inflated-toy'' transient data using an arbitrary source position and shifting the transients according to the provided scanning point locations. This unrectification simplifies adaptation of our phasor field implementation to confocal measurement geometry. In order to obtain the perturbed data, we add zero mean gaussian noise with standard deviation of 5cm along the z-axis of the provided scan locations. We then evaluate our method using the same hyperparameters as applied to our own experimental data.

We found that the amount of noise we added was sufficient to completely disrupt the phasor field reconstruction, while our method is able to produce a reasonable result. We note that the recovered reconstruction appears slightly shifted from the reconstruction obtained with no perturbation. When plotted, the recovered scan locations (Fig.~\ref{fig:scan_pos_estimate_results}) appear to converge at a plane which is at an angle from ground truth, as if the wall was slightly angled. Since we do not enforce that the scan locations are anchored to a particular orientation, our reconstruction may drift slightly with the scan locations.

\section{Discussion}
\subsection{Convergence}
Our alternating gradient update strategy performs a number of \emph{reconstruction updates} followed by \emph{calibration updates}. In the case where we only perform reconstruction updates, our reconstruction objective is convex because the forward model is linear with a mean squared error loss. Performing calibration updates in alternating fashion breaks this convexity guarantee, so we assume calibration estimates are accurate enough for gradient descent to converge to a good solution. In practice, we show good performance when refining initial calibration estimates that have enough error to break other reconstruction methods. We find that randomly sub-sampling laser scan positions ($\approx$ 10\%) and their associated measurements still lead to similar convergence performance as including all scan positions for each iteration.

When performing auto-calibration, we find that allowing laser scan positions to vary along all 3 axes leads to poor convergence to ground truth positions. Therefore, for most experiments, we only update the scanning position along the z-axis. Restricting updates to the z-axis is still useful in practice, as this direction lies approximately along the direction with most uncertainty for most experimental calibration procedures (e.g., calibration chart using calibrated cameras, or known galvo direction and unknown distance to relay wall). A possible explanation for poor performance when we update all 3 axes is as follows. For a given measured path length from a known hidden voxel position, the scanning point position may lie anywhere along the surface of an ellipsoid whose foci are the voxel position and laser source position. For most realistic L-corner geometries, the surface of the ellipsoid is approximately tangent to the relay wall surface. Thus, updates to x,y will have an insignificant impact on the loss, leading to a high condition number of the Jacobian that contributes to local instability of the gradient and poor convergence.

\subsubsection*{Recovered Calibration Parameters in Experiments}

\begin{figure}[t]
\begin{center}
  \includegraphics[width=0.9\linewidth]{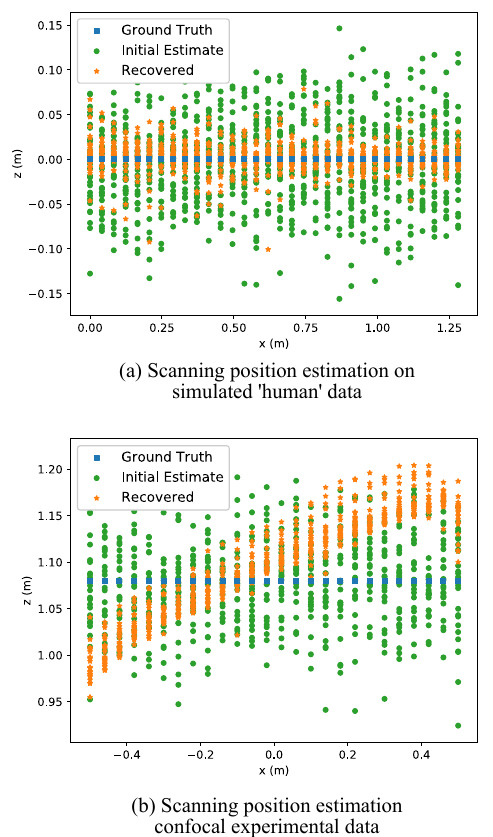}
\end{center}
  \caption{\textbf{The estimation of scanning points with the proposed automatic calibration framework.} (a) Estimate of the scanning position on the simulated ``human'' data with perturbation in z direction on the ground truth grid. (b) Automatic calibration on the confocal measurement~\cite{Ahn_2019_ICCV}.}
\label{fig:scan_pos_estimate_results}
\end{figure}

The recovered calibration parameters are shown in Fig.~\ref{fig:scan_pos_estimate_results}, projected onto the x-z plane. The recovered scan positions are lower variance than the initial estimate in both simulated and real data, suggesting convergence to better scan location prediction. For the scan locations recovered from confocal data (Fig.~\ref{fig:scan_pos_estimate_results}b), the recovered scan locations appear slightly angled from ground truth. We also tried running autocalibration on the confocal data without any artificial perturbation to the scan locations, and similar but less severe angling was observed, suggesting that the provided scan locations may be slightly miscalibrated, but given only rectified transient data we are unable to determine if miscalibration is actually present. While the provided ground truth scan locations may have a small amount of error, we believe this systematic deviation is more likely to be a result of our reconstruction converging to a local minimum.

\begin{figure}[t]
  \includegraphics[width=0.9\linewidth]{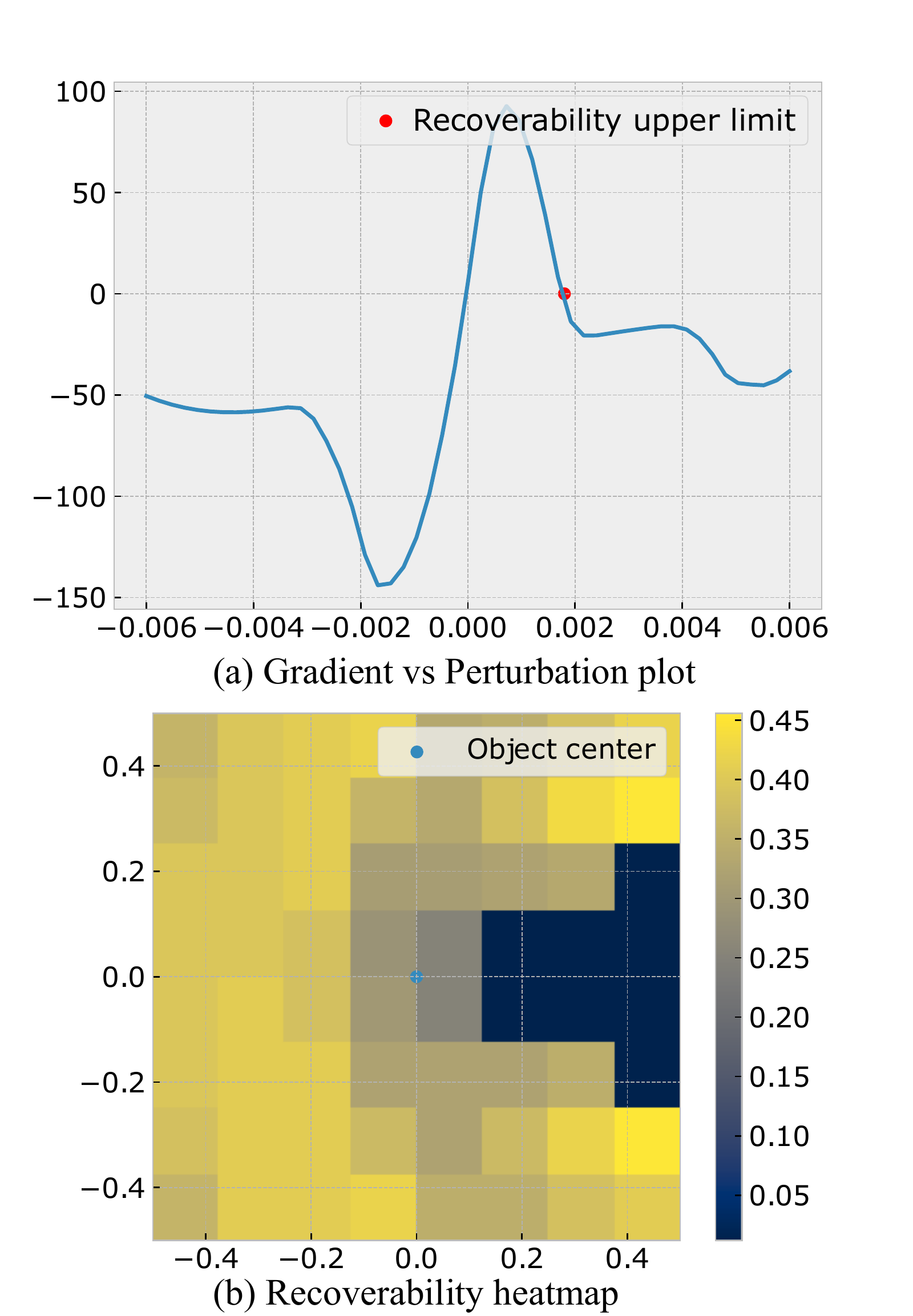}
  \caption{\textbf{Analysis of the limits of recoverability for the algorithm.}(a) Gradient vs Radial perturbation plot along transmitter-scan position. (b) Heatmap depicting the spatial distribution of the range of recoverability for the algorithm.}
\label{fig:recoverability}
\end{figure}

\subsubsection*{Simulating the Limits of Recoverability}

The algorithm has limits on the extent of perturbations that it can recover from. We show the recoverability envelope of the algorithm in Fig. \ref(fig:recoverability) b. The gradients would guide the perturbed laser scan position to its correct place within limits as positive perturbations generate positive gradients and negative perturbations generate negative giving rise to a basin of attraction. Fig. \ref(fig:recoverability) a is a heatmap showing the spatial profile of the extent of recoverability of scan positions. The color of the heatmap represents the range of perturbations over which the algorithm would be able to recover the correct scan positions at different locations. This analysis is based on simulated data with a hemispherical hidden object placed in front of the center of the scan positions, and the transmitter and the receiver are placed to the right of the scan position array.

In Fig. \ref{fig:recoverability} b, the range of recoverability reduces to the right side. We believe that this is because the transmitter is to the right and the measurements would be the most sensitive to radial perturbations to the scan positions at these locations. Fig. \ref{fig:recoverability} a shows a gradient profile which looks a lot like the gradient of a gaussian. There are other factors at play but this plot gives us an insight into the mechanism of why the algorithm works. The scale of the plot does not match the scale of the gaussian that is convolved with the simulated data. We believe that the explanation for this is the nature of the sensitivity of perturbation to the way the actual data would get affected.

\begin{figure}[t]
  \includegraphics[width=0.9\linewidth]{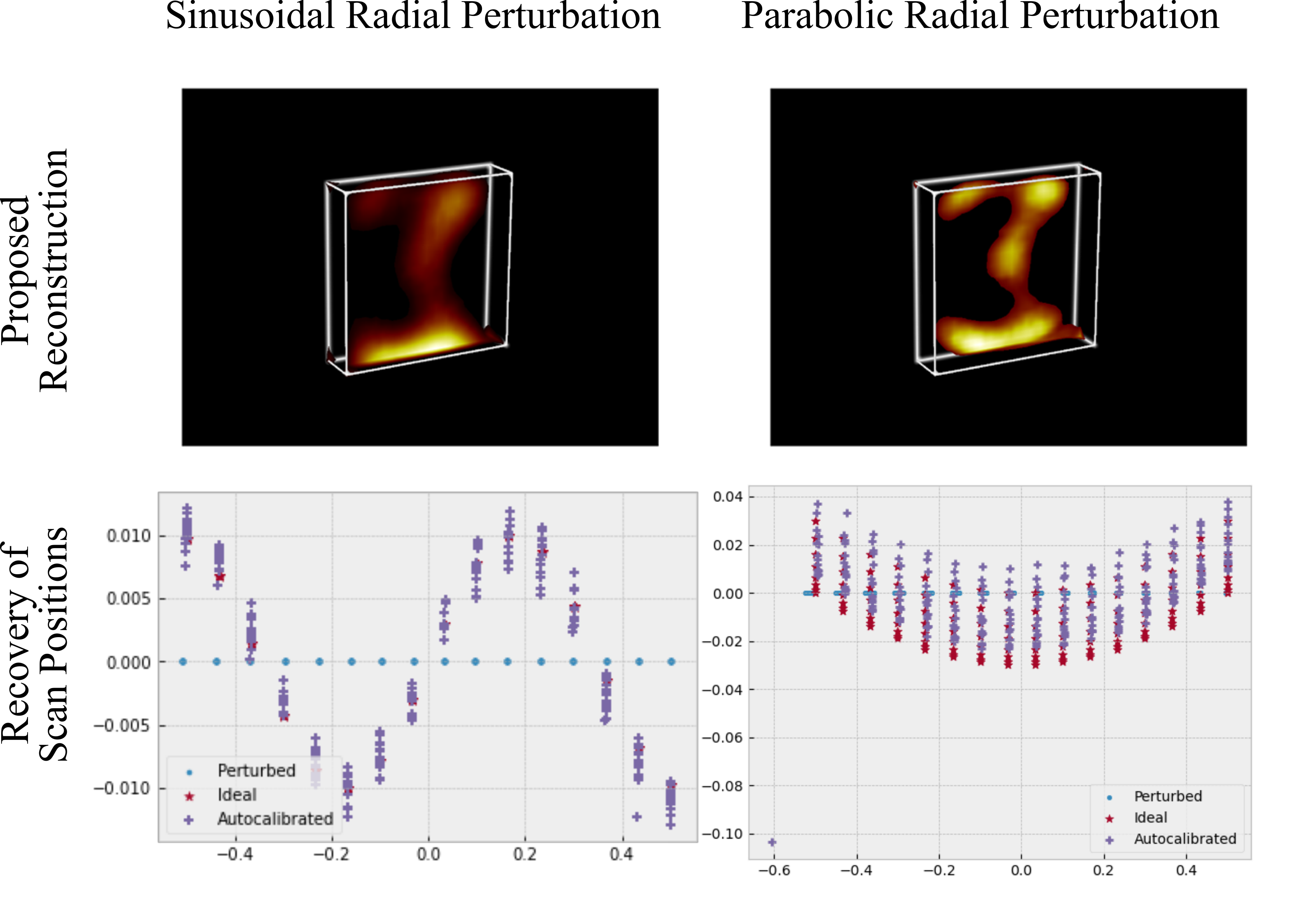}
  \caption{\textbf{Algorithm's performance across spatial patterns in perturbation.} Figure shows reconstruction of "sigma" shape with the proposed method. The scatter plot demonstrates that it is able to recover the ground truth scan positions despite varying spatial patterns.}
\label{fig:scan_pos_recovery}
\end{figure}

\subsubsection*{Effect of Spatial Patterns in Perturbation}

Fig.\ref{fig:recoverability} explains the individual behavior of the the scan positions when they go through gradient descent. In Fig. \ref{fig:scan_pos_recovery}, we demonstrate that the algorithm is capable of recovering the scan positions for not only a gaussian noisy perturbation, but also for perturbations which are of a much lower frequency. Each scan position is perturbed along the line joining the ground truth scan position and the transmitter to form a sinusoidal shape and a parabolic shape. The optimization routine is able to recover the original scan positions. We believe that this is because the gradients for scan positions are individually computed by autodifferentiation feature. So we do not see a reason for a spatial pattern to affect the optimization routine.

\subsection{Complexity}

\subsubsection{Computational complexity}
Let us assume that the total number of laser scan positions is $\mathbf{p}$, the total number of voxels is $\mathbf{q}$, the total number of time bins considered in the measurements is $\mathbf{r}$, and the total iterations is $\mathbf{i}$. The complexity of evaluating our forward model is $O(\mathbf{pqr})$. As a matrix-free method, the memory complexity of calculating our forward model is $O(\mathbf{q+pr})$, since each histogram can be calculated independently, and we only require enough memory to store the result. 

Calculating the gradients using AD complicates the evaluation of the computational complexity of our algorithm. For ``reverse-mode'' AD, such as PyTorch, the time complexity is of the same order as the evaluation of the forward model\cite{DBLP:journals/tcs/BaurS83, griewank2008}, however, gradient descent can be slow to converge, leading to an effective complexity of $O(\mathbf{ipqr})$. Our experiments typically required a few hundred iterations, where each iteration ran about as fast as computing the phasor field reconstruction. We point out that once the calibration parameters have been recovered, we can then apply a more efficient reconstruction algorithm for a given scene. 

PyTorch requires that intermediate evaluations are stored in memory for the entire forward operation. Thus memory complexity of each iteration is $O(\mathbf{pqr})$. In practice, our formulation enables easy sub-sampling of $p'$ scanning points such that memory requirements can be reduced, although doing this may change convergence behavior as the resulting gradients are only an approximation of the full gradient without subsampling.

Large memory requirements are not a fundamental limitation of AD, as gradients can be computed using ``forward-mode'' AD without changing the memory complexity of the original forward model, at the cost of quadratic time complexity. Although outside the scope of this paper, ``hybrid-mode'' AD may trade time complexity for reduced memory by re-computing only some intermediate variables, although finding the optimal computational graph is known to be NP-complete~\cite{Naumann2008}. Recent progress in AD formulations has demonstrated that good heuristics can produce reasonable trade-offs between memory and time complexity for practical problems~\cite{NimierDavidVicini2019Mitsuba2}.

\subsection{Model Mismatch on Real Data}

Our forward model assumes an underlying light transport matrix where the relay wall and hidden scene are Lambertian, no occlusion in the hidden scene volume, and an absence of noise, multiple bounces, and other sources of photon counts occurring in regions not in the reconstruction volume. Our model produces improved reconstructions of real data despite the mismatch. We observe that voxel albedo estimates at the boundaries of our reconstruction volume would continually increase in value during gradient descent. This is explained by the signal in the measured transients that extend beyond the reconstruction volume. Light transients arising from regions beyond the reconstruction volume cause unreasonable large albedo values to be calculated at the reconstruction boundaries in the process of minimizing the loss function.

We note that phasor field reconstructions performed on recovered scan locations appear to contain fewer artifacts than our own reconstructions. We believe this is the case because phasor field based reconstruction is more robust to model mismatch due to light transient measurements that arise from multipath reflections or from regions outside the reconstruction volume. Our approach complements methods such as phasor field reconstruction because once scan locations are recovered, any appropriate reconstruction algorithm can be applied.


\section{Conclusion}
In this work, we introduce the problem of auto-calibration of virtual illumination and detector positions used for time of flight based non-line-of-sight reconstruction. This is an important problem in making NLOS imaging possible without precise calibration of the visible scene geometry. We propose a modification to the traditional NLOS light transport model that ensures scanning and detector positions have non-zero gradients when implemented with an auto-differentiation framework. 

Using our modified forward model, hidden scene albedo and calibration parameters can be estimated simultaneously via gradient descent using a least squares measurement consistency loss. We validate our approach on simulated and real data and show that our method can recover from poor calibration that would otherwise result in failed reconstruction. While our method does not utilize any regularization, we expect results to improve for certain applications where scene priors exist.

Our work is the first step in taking a holistic view towards reconstruction of the hidden scene. Differentiable models such as ours open up a new direction for NLOS imaging in less controlled settings to achieve imaging ``in the wild'' where access to the visible surfaces to perform traditional calibration is infeasible. Furthermore, we think an exciting future direction would be to expand our model to accommodate estimates of other scene parameters such as material properties.

Beyond the time of flight based non-line-of-sight, making physics-based forward models differentiable with respect to parameters is an exciting frontier only recently made practical through advances and widespread adoption of automatic differentiation tools. We hope our work will inspire new robust solutions to inverse problems in imaging that are sensitive to prior calibration.

\ifpeerreview \else
\section*{Acknowledgments}
The authors would like to thank...
\fi

\bibliographystyle{IEEEtran}
\bibliography{references}

\ifpeerreview \else






\fi

\end{document}